# A GENERALIZED ANISOTROPIC DEFORMATION FORMULATION FOR GEOMATERIALS

Z. Lei, [1]E. Rougier, E. E. Knight

*Geophysics Group, Los Alamos National Laboratory, Los Alamos, New Mexico, USA 87545*

A. Munjiza

*Department of Engineering, Queen Mary, University of London, U.K, E1 4NS.*

H. Viswanathan

*Computational Earth Science Group, Los Alamos National Laboratory, Los Alamos, New Mexico, USA 87545*

ABSTRACT

In this paper, the Combined Finite-Discrete Element Method (FDEM) has been applied to analyze the deformation of anisotropic geomaterials. In the most general case geomaterials are both non-homogeneous and non-isotropic. With the aim of addressing anisotropic material problems, improved 2D FDEM formulations have been developed. These formulations feature the unified hypo-hyper elastic approach combined with a multiplicative decomposition-based selective integration for volumetric and shear deformation modes. This approach is significantly different from the co-rotational formulations typically encountered in finite element codes. Unlike the co-rotational formulation, the multiplicative decomposition-based formulation naturally decomposes deformation into translation, rotation, plastic stretches, elastic stretches, volumetric stretches, shear stretches, etc. This approach can be implemented for a whole family of finite elements from solids to shells and membranes. This novel 2D FDEM based material formulation was designed in such a way that the anisotropic properties of the solid can be specified in a cell by cell basis, therefore enabling the user to seed these anisotropic properties following any type of spatial variation, for example, following a curvilinear path. In addition, due to the selective integration, there are no problems with volumetric or shear locking with any type of finite element employed.

## 1 INTRODUCTION

Since its inception the combined finite–discrete element method (FDEM) [1-9] has become a tool of choice for a diverse field of practical engineering and scientific simulations [10-17]. Within this framework the domain discretization in 2D has been usually conducted using the constant strain triangular (CST) finite element. This element has some very advantageous features in that it can be generated easily and even automatically for complicated geometries. Moreover, the contact interaction between CSTs is relatively easy to resolve while maintaining a high computational efficiency. Although in many cases it produces very good results, "locking" problems associated with this element (in simulations that require incompressibility or near incompressibility) can seriously degrade the accuracy of the results. To overcome this problem, several approaches have been proposed, for example: smoothed finite element methods [18], nodal integration approaches [19] and composite elements [20,21]. The composite element is a good choice for FDEM especially in the cases that re-meshing is needed for the simulation of dynamic fracture propagation problems.

---

[1] Corresponding Author: Dr. Esteban Rougier, erougier@lanl.gov

The basic feature of a composite element is its "assemblage" which is composed of several sub-elements. These sub-elements are relatively independent but work together to improve accuracy of strain calculations. Camacho and Ortiz [20] described a novel triangular element in which a six-noded triangle is constructed from four three-noded triangles with linear displacement fields in each sub-triangle and a continuous linear strain field over the assemblage. Guo et al. [21] presented a detailed analysis of several composite triangular elements based on Camacho and Ortiz's work. They also proposed an alternative composite triangle in which the volumetric strain is assumed to be constant over the whole triangle.

In this work a unified constitutive approach for a 2D composite triangle has been developed. Within FDEM a large strain-large displacement formulation for the finite element side has been employed in its exact multiplicative decomposition (as opposed to co-rotational) formulation. Most recently, this formulation has been generalized through the concept of the Munjiza infinitesimal material element, which enables a pragmatic engineering approach to address anisotropic constitutive law formulations for both large displacements and large strains in the context of the exact decomposition-based format. This approach has been recently described in detail in the book entitled "Large Strain Finite Element Method: A Practical Course" by Munjiza et al. [22], where also some novel concepts of selective integration have been proposed and applied to a whole family of finite elements. One of the elements proposed in [22] is the composite triangle finite element in 2D. This element has been implemented into an in-house FDEM software package and through numerical examples, the generalized anisotropic capabilities of the unified constitutive approach have been demonstrated and are presented in this paper.

## 2 FDEM IN A NUTSHELL

The combined finite-discrete element method was proposed in the early 1990s as an alternative to describe the transition from continuum to discontinuum material behavior that occurs upon failure, i.e., fracture and fragmentation processes of brittle geomaterials [1-3]. Later on, the method was extended to cover other types of materials, such as red blood cells [4], glass [5,6], and reinforced concrete [7] among others. The use of improved finite displacements, finite rotations, and finite strain-based deformability representations combined with a variety of constitutive material laws are some of the key advantages of the FDEM. Additionally, if these features are coupled with discrete crack initiation and crack propagation models, then complex fracture patterns and fragmentation processes can be modeled.

The main parts of the FDEM can be grouped in the following four areas: a) governing equations, b) finite strain-based formulation, c) contact algorithms, and d) fracture and fragmentation solutions.

**Governing Equations.** Within the FDEM framework the solid domains are discretized into finite elements. The general governing equation that the FDEM solves is [8]:

$$\mathbf{M}\ddot{\mathbf{x}} + \mathbf{C}\dot{\mathbf{x}} = \mathbf{f} \tag{1}$$

where $\mathbf{M}$ and $\mathbf{C}$ are the lumped mass and damping matrices respectively, $\mathbf{x}$ is the displacement vector and $\mathbf{f}$ is the equivalent force vector acting on each node, which includes all forces existing in the system, such as the forces due to material deformation, contact forces between solid elements as well as cohesion forces in the damaged strain softened areas. Equation (1) is then integrated in time in order to obtain the transient evolution of the system. There are several time integration schemes that can be used for this purpose [8,24]. In this work the central difference time integration scheme was adopted.

**Finite Strain-Based Formulation.** Within the implementation of the FDEM used in this work, a multiplicative decomposition-based formulation was adopted to describe the deformation of the finite elements. This framework allows for a relatively easy transition (implementation-wise) from a hypo- to a hyper-elastic approach to describe the material deformation. A more detailed description of this point will be provided in the rest of the paper.

**Contact Algorithms.** In any standard FDEM simulation there is the need to resolve the contact physics between discrete particles. Because of this, robust and computationally efficient contact algorithms are required. In this work, the MRCK contact detection algorithm [22] was employed, along with a triangle-to-point contact interaction approach [9].

**Fracture and Fragmentation.** The single and smeared discrete crack model is used in this work [2]. In this approach actual experimental stress-strain curves are represented. The material is allowed to fail, i.e., fracture, along the finite element interfaces. The stress-strain curves at those interfaces are described by a combination of strain hardening and strain softening portions. In the strain hardening part, no failure occurs in the material and a standard continuum constitutive law is employed together with the incorporation of a non-linear elastic material description. The strain softening part of the constitutive law is represented by having stress being expressed as a function of displacement (not strain).

## 3 ELEMENT DEFORMATION FORMULATION

Within the composite triangle formulation, each six-noded triangular finite element is subdivided into four three-noded triangles, as shown in Figure 1. For each of the three-noded triangles, separate sets of local **α** and **β** material axes are introduced. The geometry of the material base at any stage during the simulation is calculated from deformation kinematics.

The starting point for this process is given by the initial position of the material axis,

$$\begin{bmatrix} \bar{\boldsymbol{\alpha}} & \bar{\boldsymbol{\beta}} \end{bmatrix} = \begin{bmatrix} \bar{\alpha}_i & \bar{\beta}_i \\ \bar{\alpha}_j & \bar{\beta}_j \end{bmatrix} \qquad (2)$$

which is provided as an input. In equation (2) the *i* and *j* sub-indexes denote the x and y components of the vectors **α** and **β**, while the "¯" on top of the variables indicates that the quantity is based on the initial coordinates of the system. In a similar manner, in the rest of this paper "^" means previous and "~" means current.

In this approach, in order to avoid the problem of volumetric locking encountered with the constant strain triangle element, selective integration is employed to allow different constitutive components to be evaluated at different integration points. Stretches generated due to deviatoric deformation are calculated at integration points $G_1$, $G_2$, $G_3$, and $G_4$ for each sub-triangle while stretches generated due to volumetric deformation are calculated for the whole composite triangle, at integration point $G_4$ (see Figure 1-a). In other words, four different infinitesimal deviatoric elements are defined for each sub-triangle, while a single infinitesimal volume element is defined for the composite triangle as a whole. The material formulation was implemented in such a way that each composite triangle can be assigned a different orientation for the material axis. This enables the simulation of complex geologic structures, where the material axes are not constant across the layers, see Figure 2.

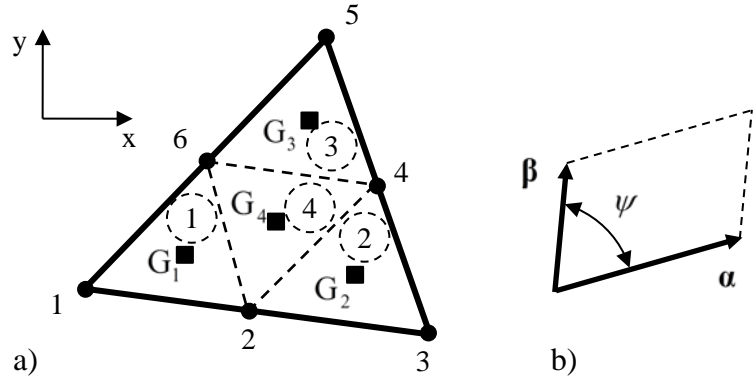

Figure 1. a) The composite triangle finite element is a six-noded triangle composed of four sub-triangles (numbers inside the dotted circles). At the center of each sub-triangle there is a material integration point ($G_1, G_2, G_3$, and $G_4$), denoted by the black squares. The node numbering order is shown next to the black dots. b) Infinitesimal solid element present at each material integration point.

For each sub-triangle the previous and the current infinitesimal solid elements, i.e., the solid elements at the previous and current time steps, are defined by

$$\begin{bmatrix} \hat{\boldsymbol{\alpha}} & \hat{\boldsymbol{\beta}} \end{bmatrix} = \begin{bmatrix} \hat{x}_a - \hat{x}_b & \hat{x}_c - \hat{x}_b \\ \hat{y}_a - \hat{y}_b & \hat{y}_c - \hat{y}_b \end{bmatrix} \left( \begin{bmatrix} \bar{x}_a - \bar{x}_b & \bar{x}_c - \bar{x}_b \\ \bar{y}_a - \bar{y}_b & \bar{y}_c - \bar{y}_b \end{bmatrix}^{-1} \begin{bmatrix} \bar{\alpha}_i & \bar{\beta}_i \\ \bar{\alpha}_j & \bar{\beta}_j \end{bmatrix} \right) \tag{3}$$

$$\begin{bmatrix} \tilde{\boldsymbol{\alpha}} & \tilde{\boldsymbol{\beta}} \end{bmatrix} = \begin{bmatrix} \tilde{x}_a - \tilde{x}_b & \tilde{x}_c - \tilde{x}_b \\ \tilde{y}_a - \tilde{y}_b & \tilde{y}_c - \tilde{y}_b \end{bmatrix} \left( \begin{bmatrix} \bar{x}_a - \bar{x}_b & \bar{x}_c - \bar{x}_b \\ \bar{y}_a - \bar{y}_b & \bar{y}_c - \bar{y}_b \end{bmatrix}^{-1} \begin{bmatrix} \bar{\alpha}_i & \bar{\beta}_i \\ \bar{\alpha}_j & \bar{\beta}_j \end{bmatrix} \right) \tag{4}$$

where the values of the indexes *a*, *b,* and *c* for each of the sub-triangles of the composite triangle are shown in Table 1.

It is worth noting that for the hyper-elastic approach the deformation of the material element is calculated by comparing the current shape (Equation (4)) versus its initial shape (Equation (2)), while for the hypo-elastic approach the deformation of the material element is calculated incrementally by comparing its current shape versus the shape it had in the previous time step (Equation (3)).

*Table 1. Indexes for each sub-triangle*

| Subtriangle | a | b | c |
|---|---|---|---|
| 1 | 2 | 1 | 6 |
| 2 | 3 | 2 | 4 |
| 3 | 5 | 4 | 6 |
| 4 | 4 | 2 | 6 |

The obtained four sets of base vectors $\begin{bmatrix} \boldsymbol{\alpha} & \boldsymbol{\beta} \end{bmatrix}$ define four different generalized material elements. These are passed to the material package (the part of the finite element code in charge of calculating the internal forces due to material deformation) together with the volumetric stretch (which is therefore the same for all four sub-triangles) resulting in a single sampling point for volumetric change and four sampling points for all the other stretches.

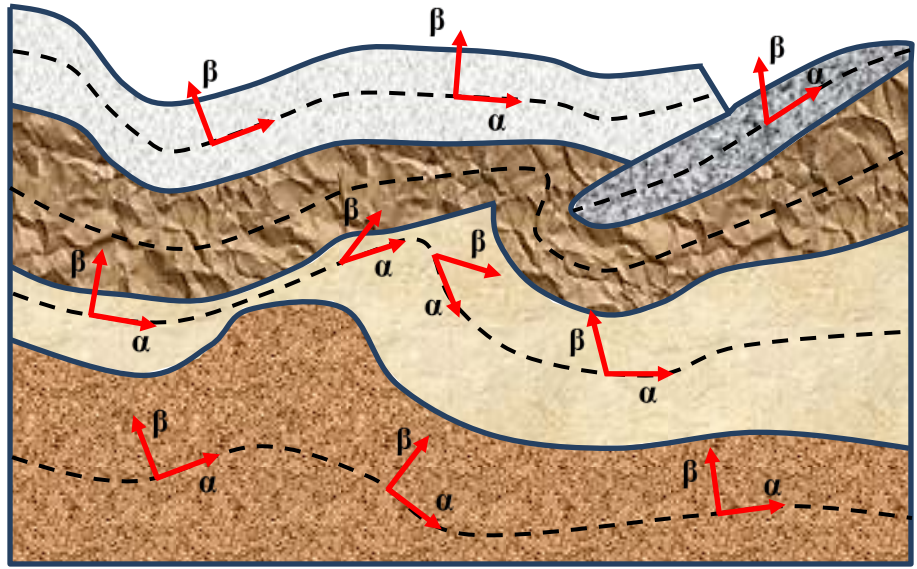

Figure 2. Orientation of material axes in a general geologic setting.

## 4 CONSTITUTIVE LAW

Very often the material law is defined in terms of a strain energy function [25]. This approach is relatively difficult to generalize to anisotropic materials. In contrast, the unified constitutive approach is relatively easy to generalize to anisotropic materials. In order to calculate the stress due to the deformation kinematics described above, the stretches of the volume, the edges, and the angle of the material element are calculated as described in the following paragraphs. In the rest of the paper only the hyper-elastic approach is described. For more details on the hypo-elastic approach the reader can refer to [22].

The volumetric stretch is obtained as follows

$$s_v = \frac{\tilde{v}}{\bar{v}} \tag{5}$$

where $\bar{v}$ is the initial volume of the generalized material element,

$$\bar{v} = \det\begin{bmatrix} \bar{\alpha}_i & \bar{\alpha}_j \\ \bar{\beta}_i & \bar{\beta}_j \end{bmatrix} = \bar{\alpha}_i \bar{\beta}_j - \bar{\alpha}_j \bar{\beta}_i \tag{6}$$

and $\tilde{v}$ is the current volume of the generalized material element,

$$\tilde{v} = \det\begin{bmatrix} \tilde{\alpha}_i & \tilde{\alpha}_j \\ \tilde{\beta}_i & \tilde{\beta}_j \end{bmatrix} = \tilde{\alpha}_i \tilde{\beta}_j - \tilde{\alpha}_j \tilde{\beta}_i \tag{7}$$

The linear stretches of the edges of the material element are given by

$$\begin{aligned} s_\alpha &= \frac{\tilde{\alpha}}{\bar{\alpha}} \\ s_\beta &= \frac{\tilde{\beta}}{\bar{\beta}} \end{aligned} \tag{8}$$

where the initial and the current lengths of the edges of the material element are given by

$$\bar{\alpha} = |\bar{\boldsymbol{\alpha}}| = \sqrt{\bar{\alpha}_i^2 + \bar{\alpha}_j^2}$$
$$\bar{\beta} = |\bar{\boldsymbol{\beta}}| = \sqrt{\bar{\beta}_i^2 + \bar{\beta}_j^2} \tag{9}$$

$$\tilde{\alpha} = |\tilde{\boldsymbol{\alpha}}| = \sqrt{\tilde{\alpha}_i^2 + \tilde{\alpha}_j^2}$$
$$\tilde{\beta} = |\tilde{\boldsymbol{\beta}}| = \sqrt{\tilde{\beta}_i^2 + \tilde{\beta}_j^2} \tag{10}$$

The angular stretch is given by

$$s_\psi = \frac{\tilde{\psi}}{\bar{\psi}} \tag{11}$$

where the initial and the current angles of the material element are given by

$$\bar{\psi} = \arccos\left(\frac{\bar{\boldsymbol{\alpha}} \cdot \bar{\boldsymbol{\beta}}}{\bar{\alpha}\bar{\beta}}\right)$$
$$\tilde{\psi} = \arccos\left(\frac{\tilde{\boldsymbol{\alpha}} \cdot \tilde{\boldsymbol{\beta}}}{\tilde{\alpha}\tilde{\beta}}\right) \tag{12}$$

The logarithmic strains are obtained from the generalized stretches as follows,

$$e_v = \ln s_v \quad ; \quad e_\alpha = \ln s_\alpha \quad ; \quad e_\beta = \ln s_\beta \quad ; \quad e_\psi = \ln s_\psi \tag{13}$$

It is worth noting that stress is a tensorial quantity (a tensor) that defines the state of internal forces (i.e., internal forces through a given surface) at a given material point P. As such, the stress tensor represents a physical reality at point P and is uniquely defined, while at the same time it can be represented by many different stress tensor's matrices. One of these is the so called Munjiza stress tensor. The components of the Munjiza stress tensor are given by

$$m_v = M_v e_v \quad ; \quad m_\alpha = M_\alpha e_\alpha \quad ; \quad m_\beta = M_\beta e_\beta \quad ; \quad m_\psi = M_\psi e_\psi \tag{14}$$

where $M_v$, $M_\alpha$, $M_\beta$, and $M_\psi$ are the Munjiza elastic constants. For the isotropic plane stress case these constants are given by

$$M_v = \frac{vE}{(1-v^2)}$$
$$M_\alpha = M_\beta = M_\psi = \frac{E}{(1+v)} \tag{15}$$

while for the isotropic plane strain case, the constants are given by

$$M_v = \frac{vE}{(1+v)(1-2v)}$$
$$M_\alpha = M_\beta = M_\psi = \frac{E}{(1+v)} \tag{16}$$

A graphical representation of each component of the Munjiza stress tensor is given in Figure 3.

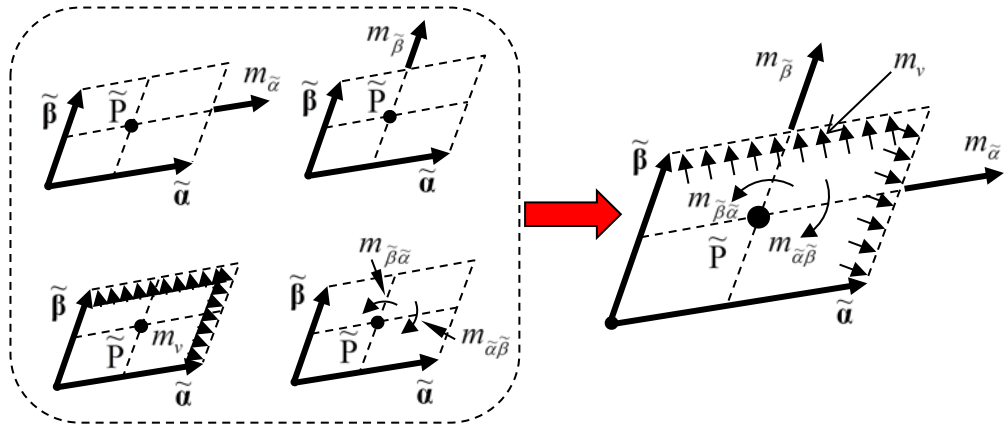

Figure 3. Components of the Munjiza stress tensor matrix.

The Munjiza stress tensor does not change with deformation (i.e., it is invariant with respect to the deformation function) and it is therefore stored within the material package – the Cauchy stress tensor is calculated from the Munjiza stress tensor matrix; the Cauchy stress tensor is not stored but is temporarily derived and used to calculate nodal forces. In this manner, any need for any familiar deformation-dependent "stress updates" (such as the Jaumann rate) is eliminated completely.

For the sake of completeness, the expression for the plane stress Munjiza elastic constants for the case of a transverse isotropic material in 2D (under the assumption of small strains) are shown below

$$M_v = \frac{\nu_{\beta\alpha} E_\alpha}{1 - \nu_{\alpha\beta}\nu_{\beta\alpha}} = \frac{\nu_{\alpha\beta} E_\beta}{1 - \nu_{\alpha\beta}\nu_{\beta\alpha}}$$

$$M_\alpha = \frac{1 - \nu_{\beta\alpha}}{1 - \nu_{\alpha\beta}\nu_{\beta\alpha}} E_\alpha$$

$$M_\beta = \frac{1 - \nu_{\alpha\beta}}{1 - \nu_{\alpha\beta}\nu_{\beta\alpha}} E_\beta$$

$$M_\psi = 2G$$

(17)

where $E_\alpha$ and $E_\beta$ are the Young's moduli for the $\alpha$ and $\beta$ directions, while $\nu_{\alpha\beta}$ and $\nu_{\beta\alpha}$ are the Poisson's ratios relating the deformations in the $\alpha \rightarrow \beta$ and $\beta \rightarrow \alpha$ sense. In this case, G is the shear modulus which is another independent parameter.

The Munjiza constants for a transverse isotropic material in 2D plane strain are given by

$$M_v = \frac{\nu_{\beta\alpha} E_\alpha}{1 - \nu_{p\alpha} - 2\nu_{\alpha\beta}\nu_{\beta\alpha}} = \frac{\nu_{\alpha\beta} E_\beta}{1 - \nu_{p\alpha} - 2\nu_{\alpha\beta}\nu_{\beta\alpha}}$$

$$M_\alpha = \frac{(1 - \nu_{\beta\alpha} - \nu_{\alpha\beta}\nu_{\beta\alpha} - \nu_{p\alpha}\nu_{\beta\alpha}) E_\alpha}{(1 + \nu_{p\alpha})(1 - \nu_{p\alpha} - 2\nu_{\alpha\beta}\nu_{\beta\alpha})}$$

$$M_\beta = \frac{(1 - \nu_{\alpha\beta} - \nu_{p\alpha}\nu_{p\alpha} - \nu_{p\alpha}\nu_{\alpha\beta}) E_\beta}{(1 + \nu_{p\alpha})(1 - \nu_{p\alpha} - 2\nu_{\alpha\beta}\nu_{\beta\alpha})}$$

$$M_\psi = 2G$$

(18)

# 5 NUMERICAL EXAMPLES

## 5.1 Wave Propagation around a Borehole

In this section, the implementation of the unified material model is illustrated using a 2D square block with a circular borehole in a center, as shown in Figure 4. The model was exercised with isotropic and with anisotropic materials that had Poisson's ratios of 0.00 and 0.25.

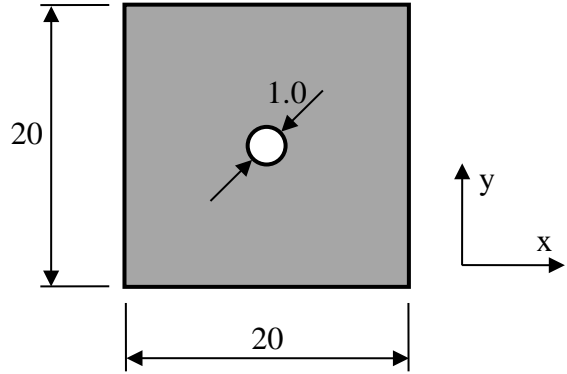

Figure 4. General dimensions of the model (all dimensions in meters).

A pressure pulse described by a Heaviside step function with a magnitude of 1.0 MPa was applied to the borehole at the beginning of the simulation. Two types of materials were considered: an isotropic material and a general anisotropic material.

For each case, four different orientations for the material axes $[\alpha \quad \beta]$ were used, as shown in Figure 5. In all four cases the material axis $\alpha$ and $\beta$ are orthogonal to each other. In Figure 5-a, the material axis $\alpha$ is collinear with the x-axis across the whole model. The material axis orientation shown in Figure 5-b is the same as the one shown in Figure 5-a, but rotated 45° counter-clockwise. Figure 5-c shows a general orientation of the material axes that changes from finite element to finite element. The orientation of the material axis shown in Figure 5-d is the same as the one shown in Figure 5-c, but rotated 45° counter-clockwise.

For the sake of comparing the waveforms obtained with the different material axes orientations, the monitoring points shown in Figure 5 were introduced. In all cases, the radial range of the monitoring points was set to 5 m. For the cases shown in Figure 5-a and Figure 5-c, the monitoring points were located along the x axis while for the cases shown in Figure 5-b and Figure 5-d their positions were rotated 45 degrees counter-clockwise to facilitate the comparison of the radial velocity histories.

**Poisson's Ratio = 0.00.** For the isotropic case the material properties were given by

$$M_v = 0.0 \, \text{GPa} \quad ; \quad M_\alpha = M_\beta = M_\psi = 1.0 \, \text{GPa} \quad ; \quad \rho = 1000.0 \, \text{kg/m}^3 \qquad (19)$$

For the anisotropic case the material properties used were the same as the ones listed above with the following exception

$$M_\beta = 2.0 \, \text{GPa} \qquad (20)$$

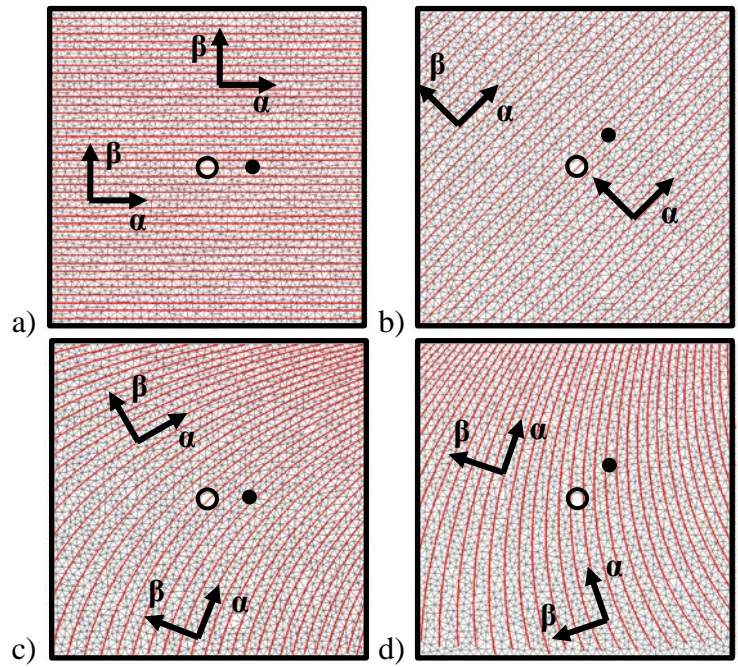

Figure 5. Orientation of the material axis **α**: a) horizontal, b) 45°, c) general #1, and d) general #2. The black dots identify the position of the monitoring points for each case.

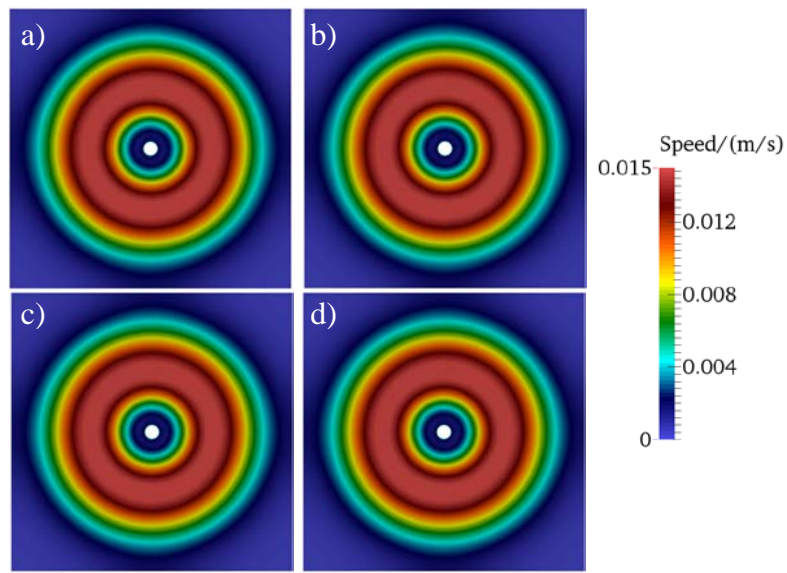

Figure 6. Wave propagation for the isotropic material model with $v = 0.00$: a) horizontal, b) 45°, c) general #1, and d) general #2.

A snapshot of the wave propagation for the case with an isotropic material is shown in Figure 6. As is expected, the wave front preserves a circular shape regardless of the orientation of the material axes. This is also confirmed by the analysis of the radial velocity time histories which are shown in Figure 7. All four curves fall on top of each other which demonstrates that the formulation is not sensitive to the orientation of the material axes.

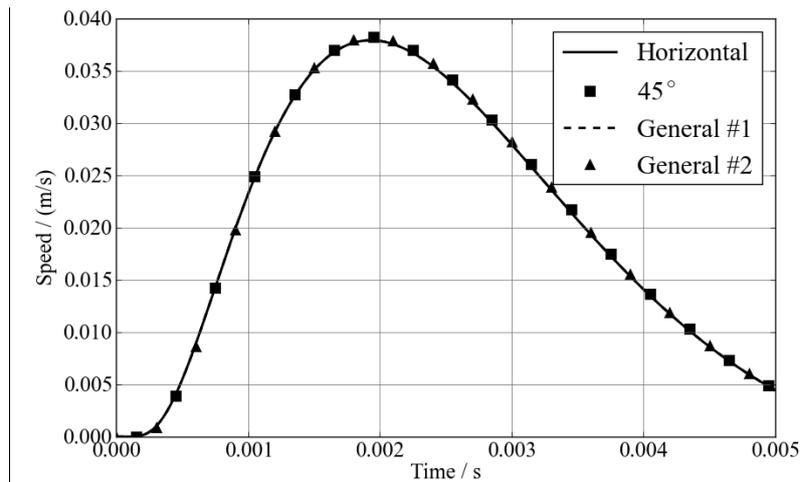

Figure 7. Isotropic Material case: comparison of the time histories of the radial speed for the monitoring points shown in Figure 5.

A snapshot of the wave propagation for the case with an anisotropic material is shown in Figure 8. In this case, the shape of the wave propagation front is strongly affected by the orientation of the material axes.

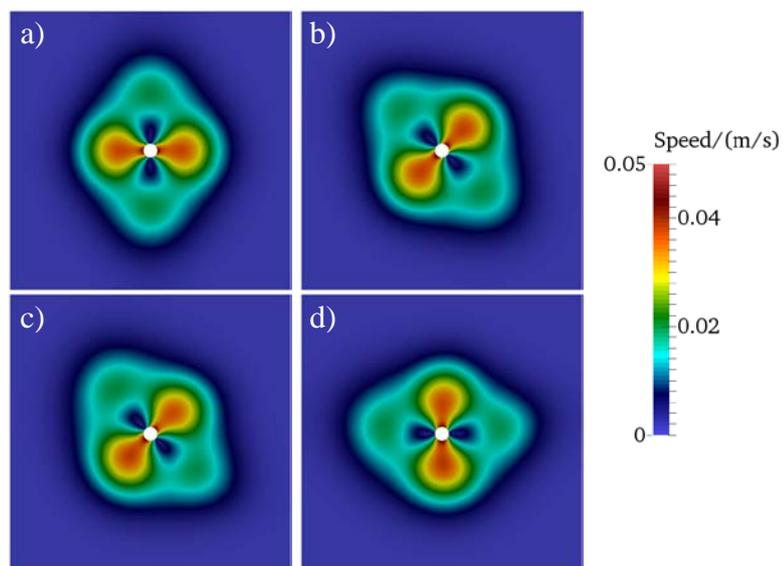

Figure 8. Wave propagation for the anisotropic material model with $v = 0.00$: a) horizontal, b) 45º, c) general #1, and d) general #2.

For the general anisotropic case the independence of the results with respect to the orientation of the material axes is demonstrated by comparing Figure 8-a against Figure 8-b rotated 45º clockwise and Figure 8-c against Figure 8-d, also rotated 45º clockwise. These comparisons are shown in Figure 9 with the help of isolines. The red isolines correspond to the un-rotated results (Figure 8-a and Figure 8-c), while the white isolines correspond to the rotated results (Figure 8-b and Figure 8-d). A comparison of the velocity histories for the two monitoring points for the general anisotropic case is shown in Figure 10. It is evident that there is no change in the wave propagation front as the material axis are rotated.

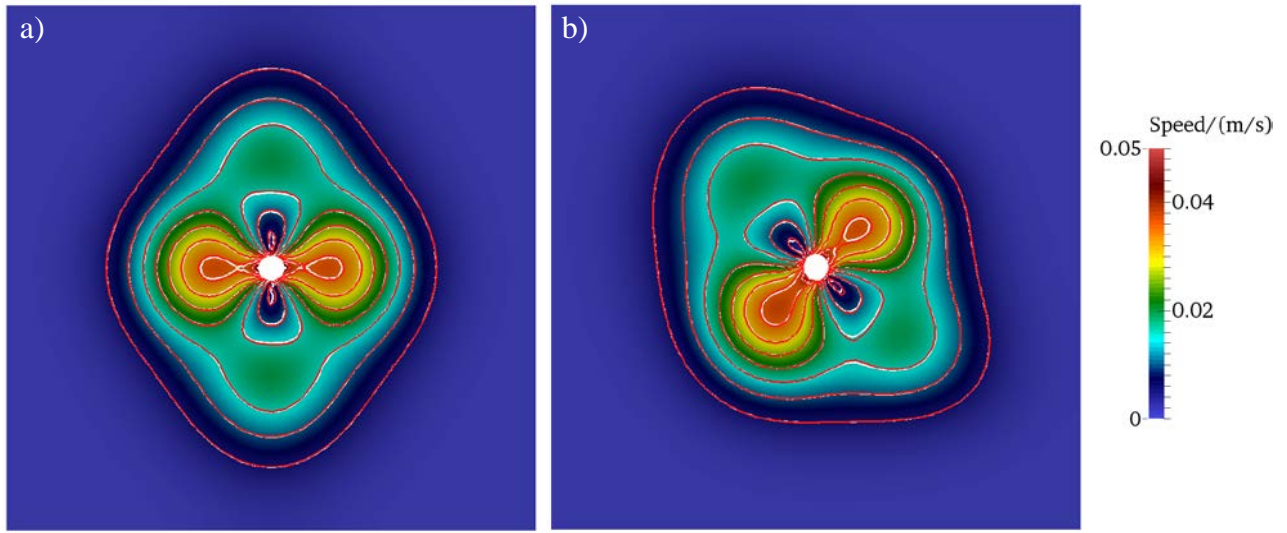

Figure 9. Wave propagation for the anisotropic material model with $\nu = 0.00$: a) horizontal and b) general.

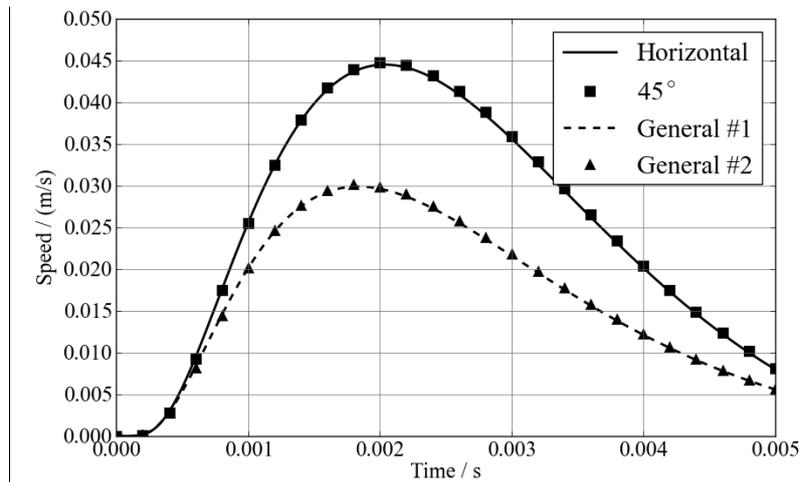

Figure 10. Comparison of radial speed for monitoring points shown in Figure 5.

**Poisson's Ratio = 0.25.** The same model used in the previous example was used to test the composite triangle formulation for a case with Poisson's ratio

$$\nu = 0.25 \tag{21}$$

For the isotropic case the material properties were given by

$$M_\nu = 0.267 \text{ GPa} \;\; ; \;\; M_\alpha = M_\beta = M_\psi = 0.800 \text{ GPa} \;\; ; \;\; \rho = 1000.0 \text{ kg/m}^3 \tag{22}$$

For the anisotropic case the material properties used were the same as the ones listed above with the following exception

$$M_\beta = 1.600 \text{ GPa} \tag{23}$$

A snapshot of the wave propagation for the case with an isotropic material is shown in Figure 11. As in the previous example, the wave front preserves a circular shape regardless of the orientation of the material axes.

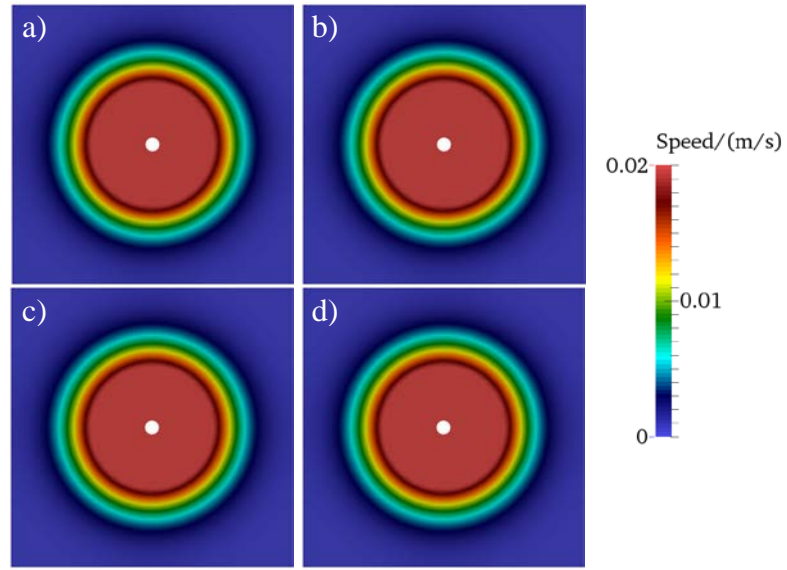

Figure 11. Wave propagation for the isotropic material model with $\nu = 0.25$ : a) horizontal, b) 45º, c) general #1, and d) general #2.

A snapshot of the wave propagation for the case with an anisotropic material is shown in Figure 12. As in the previous example, the shape of the wave propagation front is strongly affected by the orientation of the material axes.

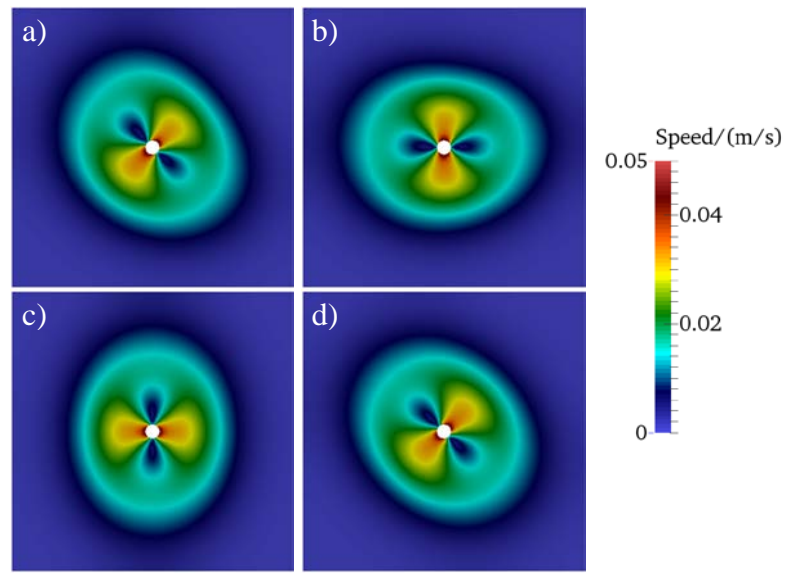

Figure 12. Wave propagation for the anisotropic material model with $\nu = 0.25$ : a) horizontal, b) 45º, c) general #1, and d) general #2.

The comparisons of the wave fronts and of the radial velocity time histories for the matching cases are shown in Figure 13 and in Figure 14. There is virtually no difference between the un-rotated and the rotated material axes cases, as it should be, as it is demonstrated by a difference of only 0.5% when comparing the corresponding peak speeds shown in Figure 14.

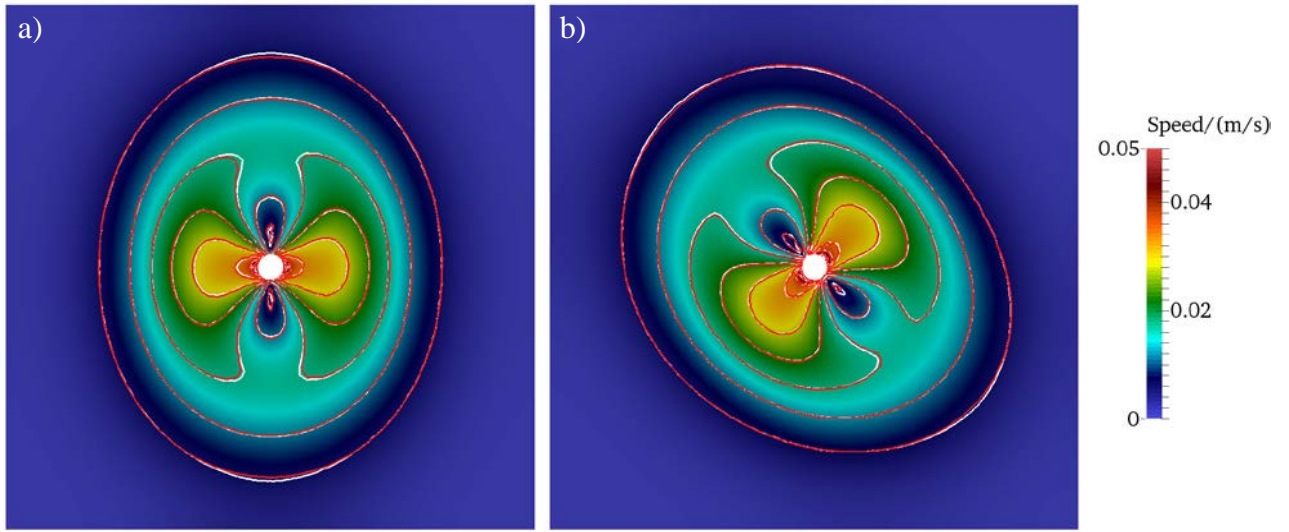

Figure 13. Wave propagation for the anisotropic material model with $\nu = 0.25$: a) horizontal and b) general.

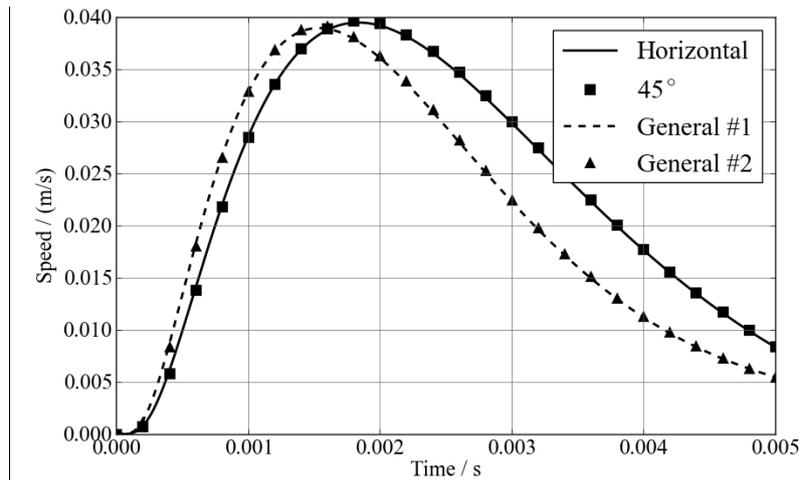

Figure 14. Comparison of radial speed for monitoring points shown in Figure 5.

## 5.2 Behavior under Volumetric Locking Conditions

Figure 15 shows one of the models that was used to demonstrate that the composite element proposed here does not suffer from the problems generated by volumetric locking. The model consists of a square with the left and bottom edges fixed and a force

$$\mathbf{f} = \begin{bmatrix} f_i \\ f_j \end{bmatrix} \quad (24)$$

applied to the corner node. The example was run under plane stress and plane strain conditions utilizing a standard constant strain triangle (CST) formulation and the composite triangle (COMPT) proposed in this work. Since the aim of this exercise was to obtain the quasi-static response of the system, the simulations were damped accordingly. Therefore, the results shown in this section correspond to the static solutions.

For the plane stress case the Munjiza elastic constants used are

$$M_v = M_\alpha = M_\beta = M_\psi = 0.667 \text{ GPa} \quad ; \quad \rho = 1000.0 \text{ kg/m}^3 \quad (25)$$

These constants were derived using Equation (15) considering $E = 1.0 \text{ GPa}$ and $\nu = 0.5$. For the plane strain case the Munjiza elastic constants used are

$$M_v = 1,666.4 \text{ GPa} \quad ; \quad M_\alpha = M_\beta = M_\psi = 0.667 \text{ GPa} \quad ; \quad \rho = 1000.0 \text{ kg/m}^3 \quad (26)$$

It is worth noting that in this case the Poisson's ratio was set to 0.4999.

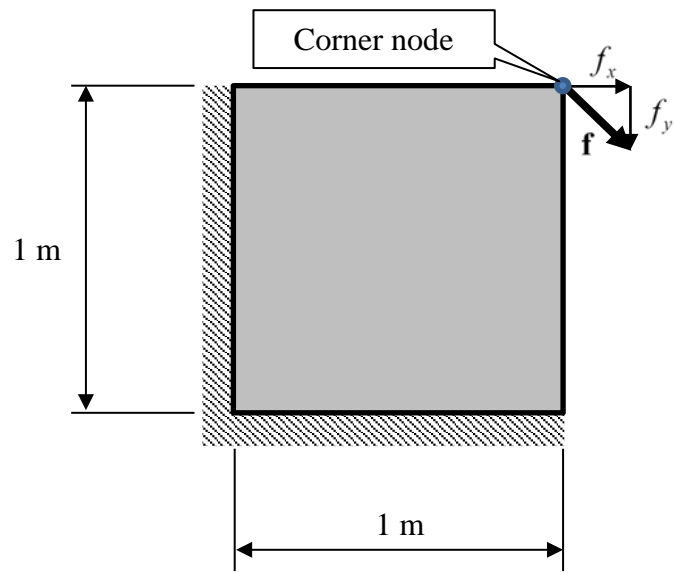

Figure 15. Model setup for volumetric locking test with corner force.

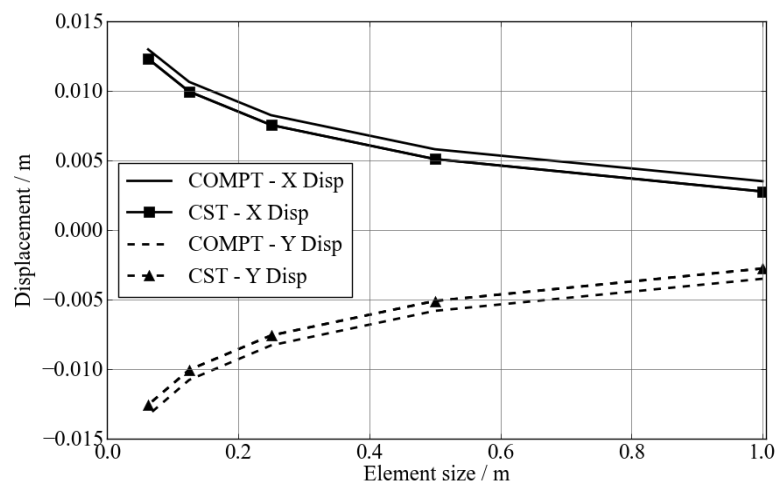

Figure 16. Evolution of the x- and y-displacements of the corner node for the plane stress case.

The results obtained for the x- and y-displacements of the corner node as a function of the element size for the plane stress case are shown in Figure 16. The response obtained from the COMPT model is consistently softer that the one obtained from the CST model, as it is expected to be. These differences in displacement response do not change substantially relative to each other as the element size is decreased. A snapshot of the x component of the Cauchy stress tensor ($\mathbf{C}_{xx}$) for both cases is shown in Figure 17. The stress map in both cases is quite uniform, except around the corner where the force is being applied.

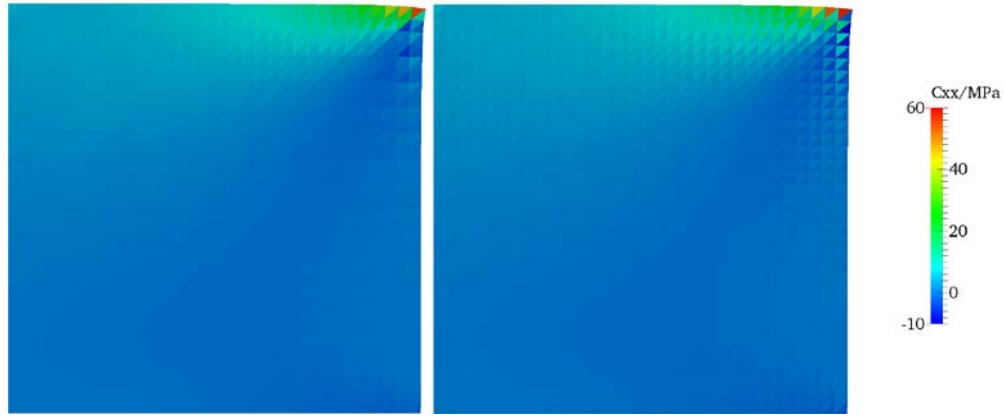

Figure 17. Stress distribution in the plane stress case: a) with the COMPT and b) with the CST.

For the plane strain case, the story is quite different. The x- and y-displacements of the corner nodes as a function of the element size are shown in Figure 18. The CST case exhibits very strong volumetric locking effects resulting in an almost non-existent displacement of the corner node. On the other hand, the COMPT model is not affected by this, as is demonstrated by the curves shown in Figure 18.

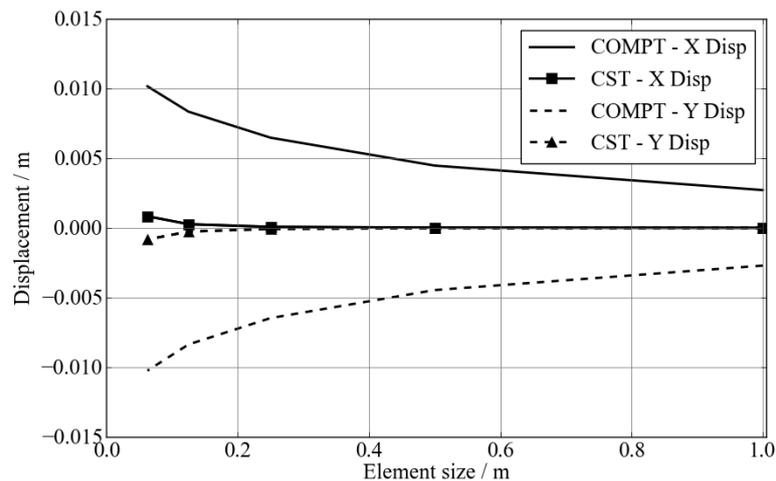

Figure 18. Evolution of the x- and y-displacements of the corner node for the plane strain case.

The snapshots of the x component of the Cauchy stress tensor for the CST and for the COMPT models in plane strain conditions are shown in Figure 19. The stress field obtained for the CST shows a very strong "checkerboard" pattern effect caused by the volumetric locking of the elements. On the other hand, the stress field obtained for the COMPT is very smooth, with only some incipient evidence of "checkerboard" pattern effect close to the corner where the force is being applied.

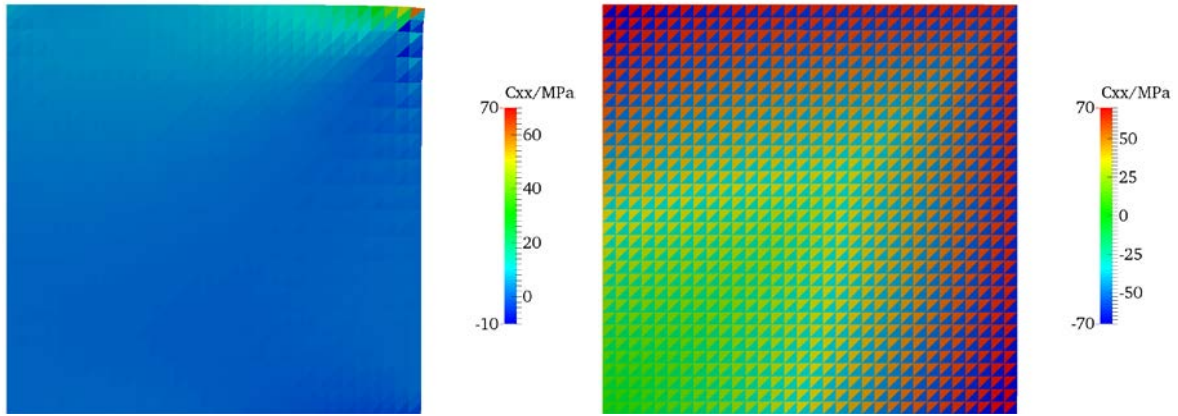

Figure 19. Stress distribution in the plane strain case: a) with the COMPT and b) with the CST.

Another example to demonstrate the absence of volumetric locking problems for the composite triangles is shown in Figure 20. In this case, a constant body force $p = 1000$ Pa was applied to both sides of the sample, as shown in the figure. In this example, two different values of the Poisson's ratio were tested under plain strain conditions: $\nu_1 = 0.25$ and $\nu_2 = 0.4999$. The results obtained for both cases are shown in Figure 21. For the cases of $\nu = 0.25$ there is no significant difference between the results obtained by using either the constant strain triangle or the composite triangle. However, when $\nu = 0.4999$ the consequences of the constant strain triangle locking problems are quite evident, as demonstrated by the checkerboard pattern effect shown in Figure 21-d.

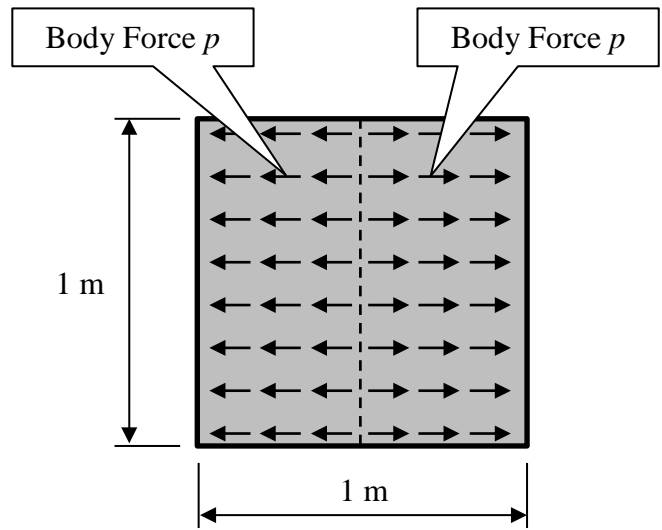

Figure 20. System setup for volumetric locking test with body force.

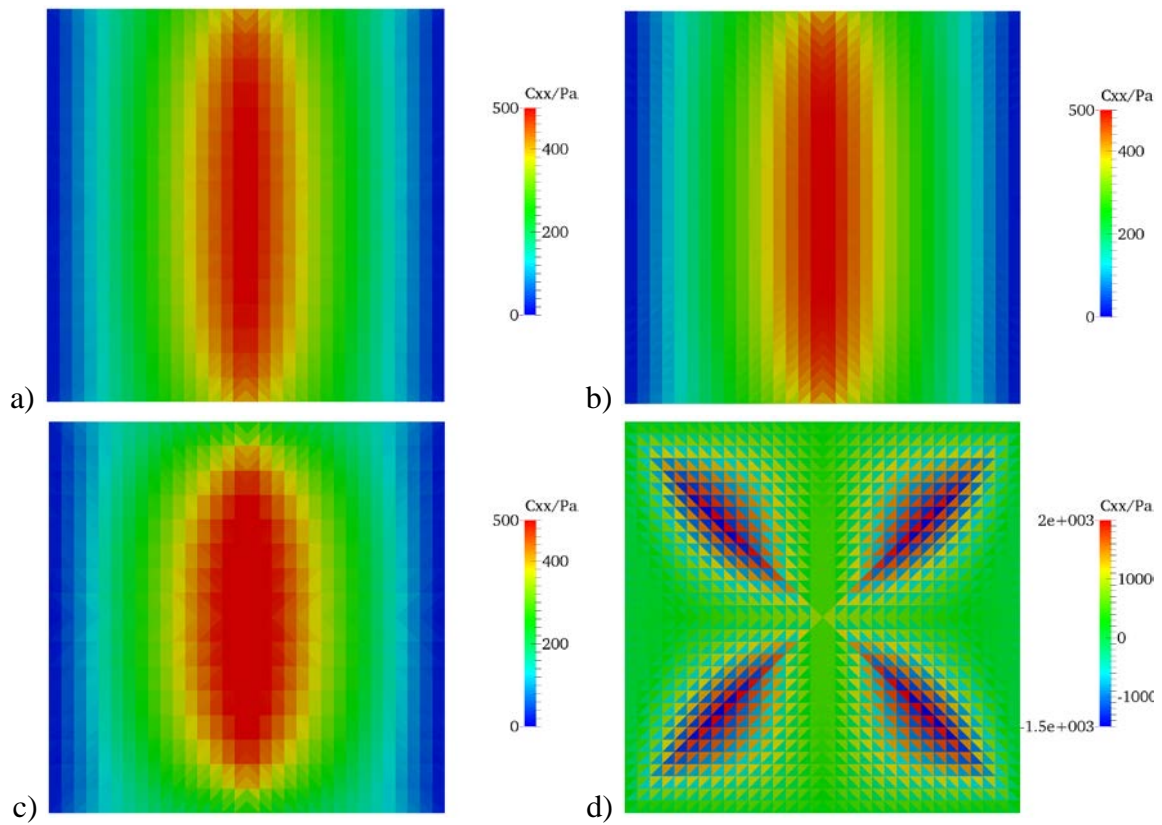

Figure 21. Stress distribution in the plane strain case: a) COMPT – $\nu_1 = 0.25$, b) CST – $\nu_1 = 0.25$, c) COMPT – $\nu_2 = 0.4999$, and d) CST – $\nu_2 = 0.4999$.

### 5.3 Numerical Illustration: Influence of Anistropic Material Properties on Geomaterials under Blast Loading

Finally, a more realistic type of problem was simulated using both isotropic and anisotropic material descriptions. For this purpose, the model shown in Figure 22 was utilized.

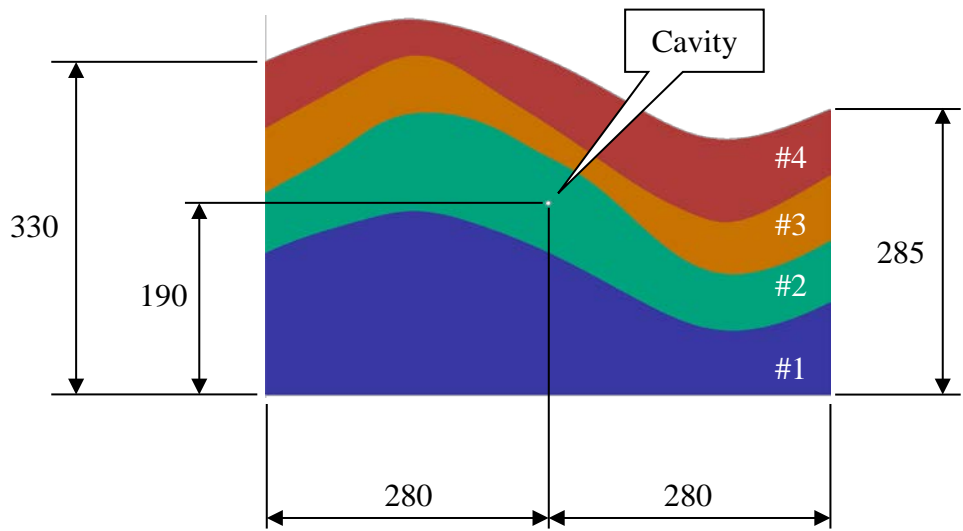

Figure 22. General description of the model of a representative geologic structure (all dimensions in meters). The radius of the cavity is 2.5 meters. Four different types of materials are included in the model.

The model consists of a section of a generic geological structure containing four different layers of different materials. For both the isotropic and the anisotropic cases, the material properties of each of the layers are listed in Table 2 and Table 3 respectively.

*Table 2. Material properties for each layer of material shown in Figure 22. Isotropic case. All values in GPa.*

| Layer # | $M_v$ | $M_\alpha$ | $M_\beta$ | $M_\psi$ |
|---|---|---|---|---|
| 1 | 6.7 | 20.0 | 20.0 | 20.0 |
| 2 | 5.3 | 16.0 | 16.0 | 16.0 |
| 3 | 8.0 | 24.0 | 24.0 | 24.0 |
| 4 | 6.7 | 20.0 | 20.0 | 20.0 |

*Table 3. Material properties for each layer of material shown in Figure 22. Anisotropic case. All values in GPa. See Figure 23-a.*

| Layer # | $M_v$ | $M_\alpha$ | $M_\beta$ | $M_\psi$ |
|---|---|---|---|---|
| 1 | 6.7 | 20.0 | 6.7 | 13.3 |
| 2 | 5.3 | 16.0 | 6.7 | 11.3 |
| 3 | 8.0 | 24.0 | 8.0 | 16.0 |
| 4 | 6.7 | 20.0 | 5.3 | 12.7 |

The cavity located inside of the layer of material #2 is subjected to a pressure boundary condition with a time history as shown in Figure 23-b. The sequence of the wave propagation for both types of mediums, i.e., anisotropic and isotropic are shown in Figure 24 and Figure 25. In the numerical results presented, the fractures are plotted in white while the wave propagation is colored according to the particle speed.

Figure 24 and Figure 25 clearly show the effects of the material descriptions on both the wave and fracture propagation. In the anisotropic case the shape of the wave propagation front is strongly affected by the orientation of the material axes, which further influences fracture initiation and propagation as well as the final patterns. This clearly demonstrates that modelers should always account for anisotropic material properties if the geophysical characterization shows that the substructure contains highly non-homogenous non-isotropic material. The above anisotropic deformation approach is one potential choice that can be utilized to handle these type of problems.

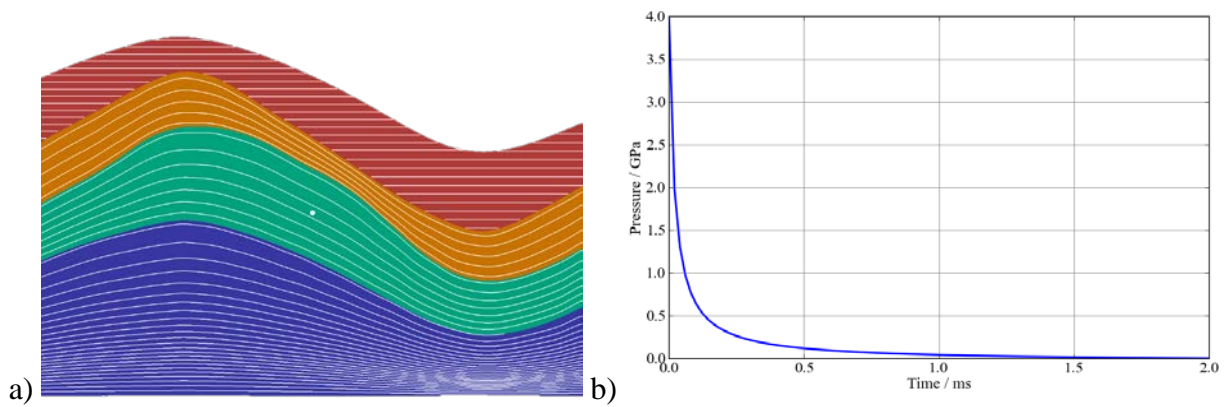

Figure 23. a) General description of the anisotropic model of a representative geologic structure. The white lines represent the direction of the material axis α with material axis β being perpendicular to α for the whole model, and b) Time history of the pressure pulse applied to the circular cavity.

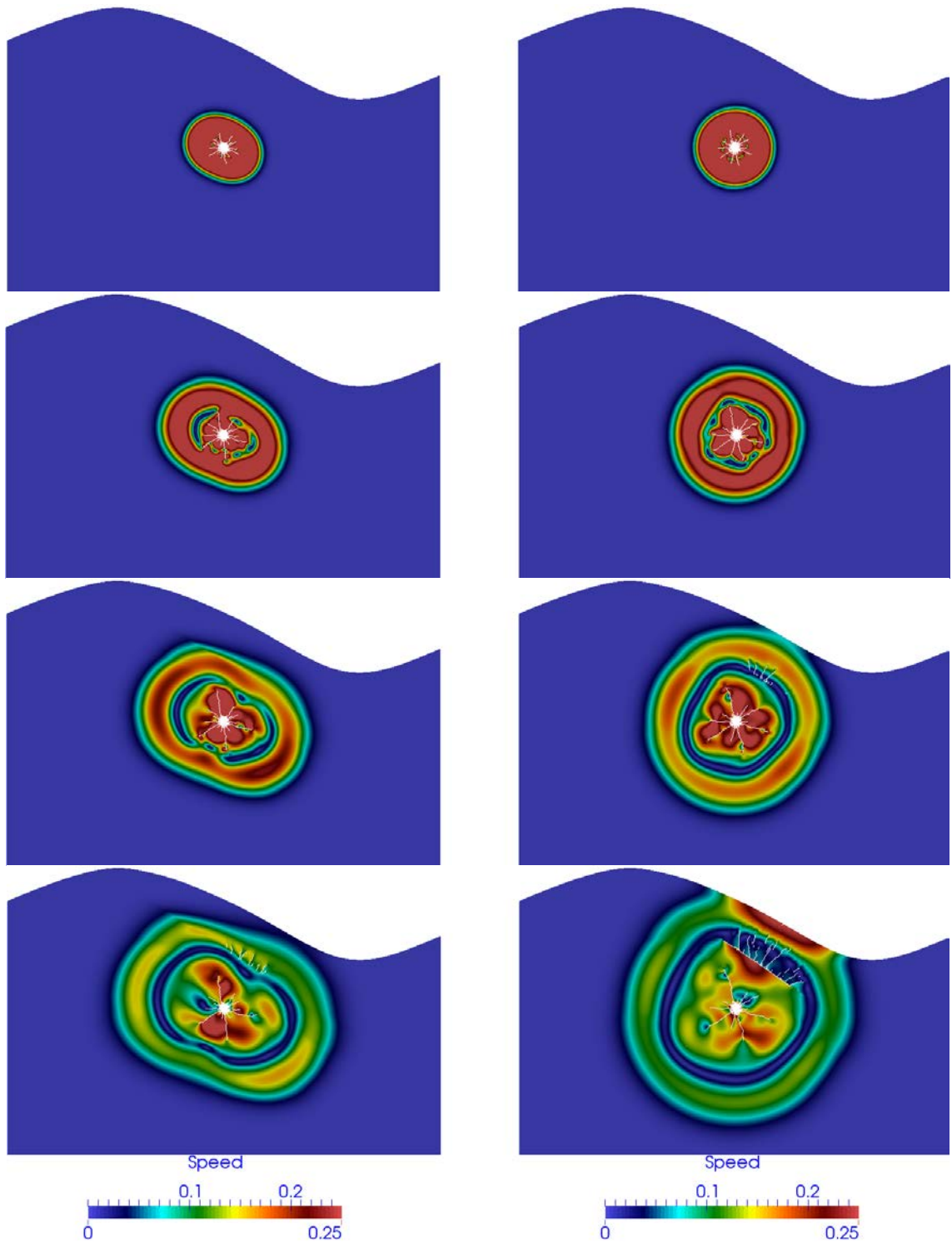

Figure 24. Comparison of the initial stages of the wave propagation through the geologic medium: Left: anisotropic material and Right: isotropic material.

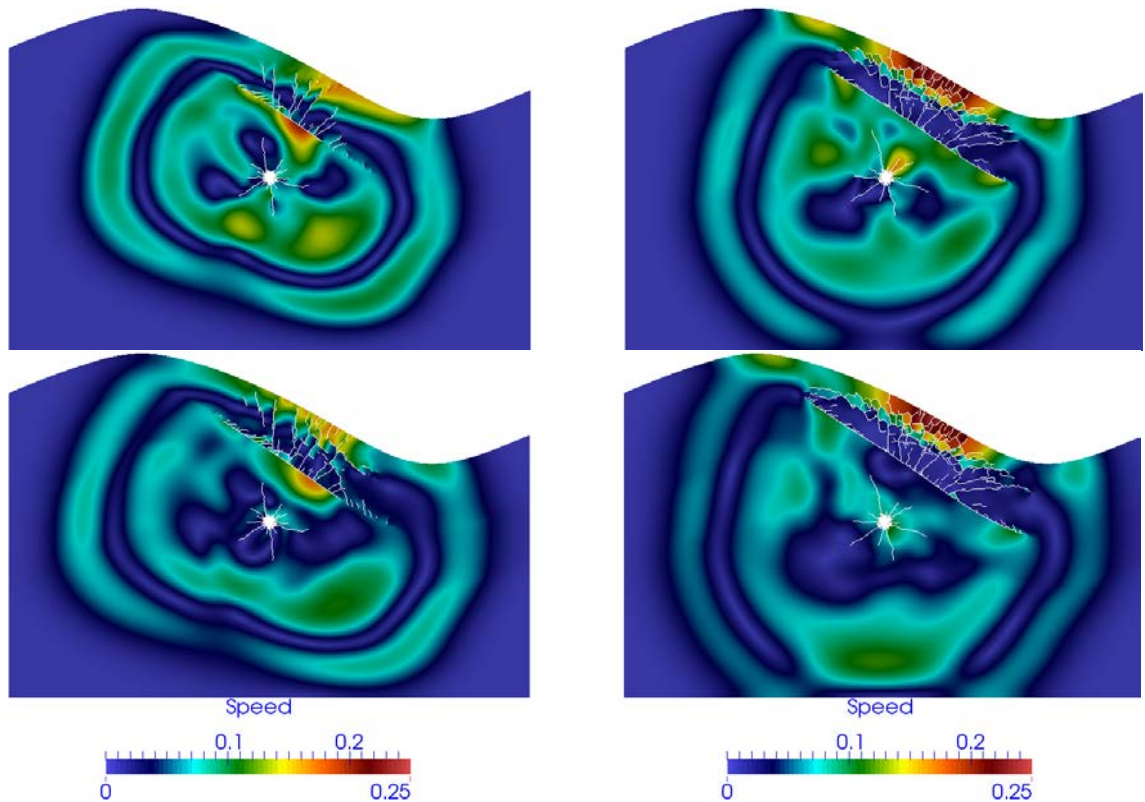

Figure 25. Comparison of the final stages of the wave propagation through the geologic medium: Left: anisotropic material and Right: isotropic material.

## 6     CONCLUSIONS

Recently published generalized deformation kinematics opens a whole range of possibilities for developing material laws in the context of large-strain large-displacement formulations for solid deformation - instead of using traditional approaches to defining material laws such as energy functions, a pragmatic engineering approach for calculating stress from stretches has been described in detail in [22].

It is worth mentioning that this represents, in a sense, a whole family of approaches that can lead to deformation independent stress tensor matrices. One of these matrices is the Munjiza stress tensor matrix.

In this work, a very brief introduction to the FDEM has been provided, along with a succinct description of the generalized deformation kinematics approach for a composite triangle in 2D. A detailed description of the hyper-elastic constitutive law for plane stress under the generalized approach has been presented. A number of well documented numerical examples were presented, where it was clearly demonstrated that: a) the results obtained with the formulation described in this work are invariant under rotation of the axes; b) the approach allows for the description of general anisotropic maps, with arbitrary spatial variations; c) the composite triangle proposed does not suffer from volumetric locking issues; and d) this approach is well suited to describing anisotropic layered geomaterials to the point that a whole tensor field of material orientations is supplied to the input - leading to different stress wave patterns.

*This is a pre-print of an article published in Computational Particle Mechanics. The final authenticated version is available online at:* https://doi.org/10.1007/s40571-015-0079-y

ACKNOWLEGDMENTS

We thank the Los Alamos National Laboratory LDRD Program (#200140002DR) for the financial support.